\documentclass[aps,prc,twocolumn]{revtex4}
\usepackage{graphicx,epsfig}% Include figure files
\usepackage{bm}% bold math

\begin{document}
\vspace{1cm}
\title{DILEPTON PRODUCTION IN NON-EQUILIBRITED QUARK-GLUON PLASMA.}
\author{S.S.Singh$^1$ and  Agam K. Jha$^1$}
\email{sssingh@physics.du.ac.in,akjha@physics.du.ac.in.}
\affiliation{ Department of Physics, University of Delhi, Delhi - 110007,INDIA}
\begin{abstract}
  A model of cut-off momentum distribution functions in quark-gluon plasma with finite baryon chemical potential is discussed.This produces
 a quark-gluon plasma signature in relativistic nuclear 
collisions with the specific structure of the dilepton spectrum 
in the transverse momentum region of (1-4) GeV and the dilepton 
production rate is found to be a strong decreasing function of chemical potential.
\end {abstract}
%\pacs{PACS number(s): 25.75.Ld, 12.38.Mh, 21.65.+f}
\maketitle
 Quantum chromodynamics(QCD), the fundamental theory of strong 
interactions\cite{wilczek}, predicts a phase transition from the state of 
hadronic constituents to a phase of deconfined quarks and gluon,so called 
Quark Gluon Plasma (QGP). Presumably the early universe was in this state 
up to about $10{\mu}$s after the big bang. Today the core program of 
ultrarelativistic nucleus-nucleus collision is to study the properties 
of the strongly interacting matter at very high energy density and 
temperature so that it can create this deconfined state of matter 
in heavy ion collision experiments. This new state of matter emits a large 
number of particles. Among these particles,electromagnetic 
probes which 
are known as photons leave this plasma volume without further 
rescattering 
and it carries the whole information about the existence of the plasma.
This produced virtual photons are subsequently decayed into dileptons 
such as $l^{-}l^{+},\mu^{+}\mu^{-}$ and they represent as a possible signature 
for the formation of QGP.
\par
  So far, many authors have studied the production of dilepton in a QGP 
at finite temperature and this temperature is related to the 
energy density $\epsilon$ given by the stefan boltzmann
$\epsilon~\sim ~ T^{4}$. However experiments at AGS and SPS energies\cite{nagamiya}
indicate that there are a sizable amount of chemical potential and 
even  at RHIC energies $\sqrt{s} \leq 200\, A GeV$,this chemical potential $\mu$ may not be neglected and calculation using microscopic 
models\cite{gustafson,mohring} limit that the colliding heavy ions may not 
be fully transparent. In this case the dilepton production rate 
is in (local) thermodynamic equilibrium ,is a function
of both temperature and quark chemical potential $\mu$ of the 
QGP which is recently studied by Dumitru et. al.\cite{dumitru}.
Again, recent work by Hammon and coworkers\cite{hammon} have indicated that the 
initial QGP system produced at the RHIC energies have finite 
baryon density. Mujumder et. al.\cite{majumdar} have discussed 
the dileptons from QGP produced at the RHIC energies at 
finite baryon density  and Bass et. al. \cite{bass} have pointed out that 
the parton rescattering  and fragmentation lead to  a substantial 
increase in the net -baryon density 
at midrapidity region. Now we consider this influencing quark chemical 
potential for one dilepon production. 
As stated above the Stefan- Bolzmanm relation $\epsilon \sim T^{4}$ is free 
from the relation $\frac{\mu}{T}$. 
We use EOS for the plasma which relates $\epsilon$,T and $\mu$.
We consider the grand partition function for 
fermion (particle and anti particle ) and boson for finding EOS 
of relativistic hydrodynamical model.Thermodynamic potentials for fermion
 and  boson are defined in rich baryon QGP as:
\begin{eqnarray}
   (Tlnz)_{f}&=&\left(\frac{g_{f}V}{12}\right) \nonumber\\
       &\times& \left(\frac{7\pi^{2}}{30}T^{4}+\mu^{2}T^{2}+ 
           \frac{1}{2\pi^{2}}\mu^{4} \right) \nonumber\\
\\   
(Tlnz)_{b}&=&\frac{g_{b}V\pi^{2}}{90}T^{4}
\end{eqnarray}
where $g_{f}$and $g_{b}$ are the respective degeneracies that is degree of freedom. The energy density  
and number density are obtained through this grand partition function.They are
$n = \frac{1}{V}\partial{(T\,lnZ)}/{\partial{\mu}}$, $\epsilon=T\partial{(T\, lnZ)}/{\partial{T}}+\mu n$
 and  we can easily establish the equation of state (EOS)
of the ideal gas $p=\frac{\epsilon}{3}$ but this QGP is far from the ideal gas at temperature as high as several times $T_{c}$
In this paper we study the dilepton production of the expanding  hot 
baryon-rich QGP on the basis  of cut off momemtum distribution function 
which are used  by Gorenstein et. al.\cite{gorenstein} in the calculation 
of dilepton production  in small invariant mass system in free baryonic system.
\par
{\bf Dilepton production from QGP}: There are a good number of
research works in dilepton production into the quark-gluon plasms. 
Whenever there is a new model for plasma evolution then its impact on
dilepton and virtual photon production are assessed. The dominant
reaction for thermal emission of dilepton pairs\cite{ruuskanen} is the Drell-Yan
mechanism $q\bar{q}\rightarrow l^{+}l^{-}$ or
$q(\bar{q})g \rightarrow q(\bar{q})+ l^{+}l^{-}$. But, we 
exclusively use  $qg\rightarrow l^{+}l^{-}$  reaction. 
The dilepton production rate $Y=\frac{dR}{d^{4}P}$(i.e the number of 
dilepton produced per space time volume  and four dimentional 
momentum space volume is given by
\begin{eqnarray}
\frac{dR}{d^{4}P}&=&\int \frac{d^{3}k_{1}}{(2 \pi)^{3}} 
\frac{d^{3}k_{2}}{(2 \pi)^{3}} f_{q}(k_{1},T,\mu) 
f_{g}(k_{2},T,\mu)\nonumber 
\\
&\times &v_{q\bar{q}} \sigma_{q\bar{q}}(M^{2}) \delta^{4}(P-k_{1}-k_{2})
\end{eqnarray}
where, 
\begin{equation}
 f_{q}(k_{1},T,\mu)=\frac{d_{q} 
\Theta(|k|-K_{q})}{\exp^\frac{(E-\mu)}{T}+1},
f_{\bar{q}}(k_{2},T,\mu)=\frac{d_{q}
\Theta(|k|-K_{q})}{\exp^\frac{(E+\mu)}{T}+1}
\end{equation}
at equilibrium, these two distribution functions are equal
\begin{equation}
f_{q}(k_{1},T,\mu=0)=f_{\bar{q}}(k_{2},T,\mu=0)
\end{equation}
and
\begin{equation}
f_{g}(k,T,\mu)=\frac{d_{g}\Theta(|k|-K_{g})}{\exp^\frac{E_{g}}{T}-1}
\end{equation}
where $d_{g}$ and $d_{q}$ are the degeneracy factor  for gluon and quarks. 
$v_{q\bar{q}}$ is the relative velocity between quark and 
antiquark. $p_{\mu}$ is lepton pair four momentum.($M^{2}=p^{\mu}p_{\mu}$ 
invariant lepton mass ).$\sigma_{{q\bar{q}}
{\rightarrow}{l\bar{l}}}$ is the electromagnetic 
annihilation cross section.It is given:
\begin{eqnarray}
\sigma_{q\bar{q}\rightarrow l^{-}l^{+}}(M^{2})
&=&16\pi \alpha^{2} \sum_{f=1}^{2}
(\frac{e_{f}}{e})^{2} \frac{1}{M^{2}} \nonumber \\
& &(1+\frac{2 m^{2}}{M^{2}})
(1-\frac{4 m^{2}}{M^{2}}).
\end{eqnarray}
where $(\frac{e_{f}}{e})^{2} =\frac{5}{9}$ the electric charge of the quark 
in the units of the electron charge, $\alpha=\frac{1}{137}$ and $m$ is the 
lepton mass but we consider the lepton mass is zero(ie $m=0$). 
Substituting  in the  equation  (3) using (4)and (6) we integrate 
over $q$ and $\bar{q}$ momentum and obtain the dilepton production rate as:
\begin{equation}
\\
 \frac{dR}{d^{4}P} = \frac{5 \alpha^{2}}{18 \pi^{3}}
\frac{T \Theta(|E|-2K_{q})}{ (\exp^{\frac{E}{T}}-1)}
ln(1-\exp^{-(E+\mu)/T})\sqrt{\frac{\pi M T}{2}} 
\end{equation}
%%%%%%%%%%%%%%%%%%%%%%%%%%%%%%%%%%%%%%%%%%%%%%%%%%%%%%%%%%%%%%%%%%%%%%%%%
\begin{figure}
\resizebox*{3.1in}{3.1in}{\rotatebox{270}{\includegraphics{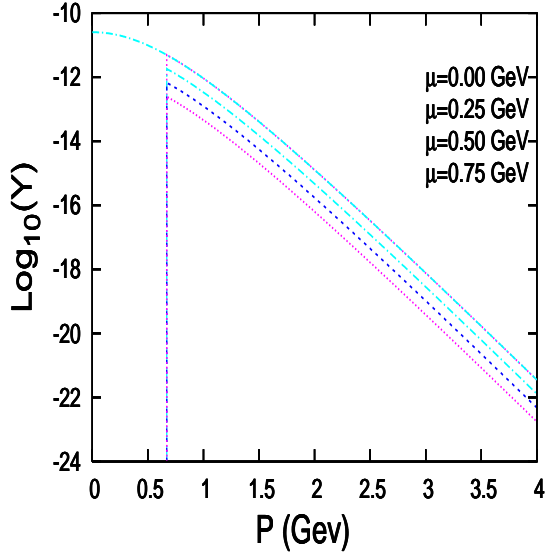}}}
\vspace*{0.5cm}
\caption[]{
The dilepton emission rate,$Y=\frac{dR}{d^{4}P}$,
log($Y$) at parameter 
$M=1.0 GeV, T=0.25 GeV,K_{q}=0.6 GeV$ for different values of$\mu$}
\label{scaling}
\end{figure}
%%%%%%%%%%%%%%%%%%%%%%%%%%%%%%%%%%%%%%%%%%%%%%%%%%%%%%%%%%%%%%%%%%%%%%%%
\begin{figure}
\resizebox*{3.1in}{3.1in}{\rotatebox{270}{\includegraphics{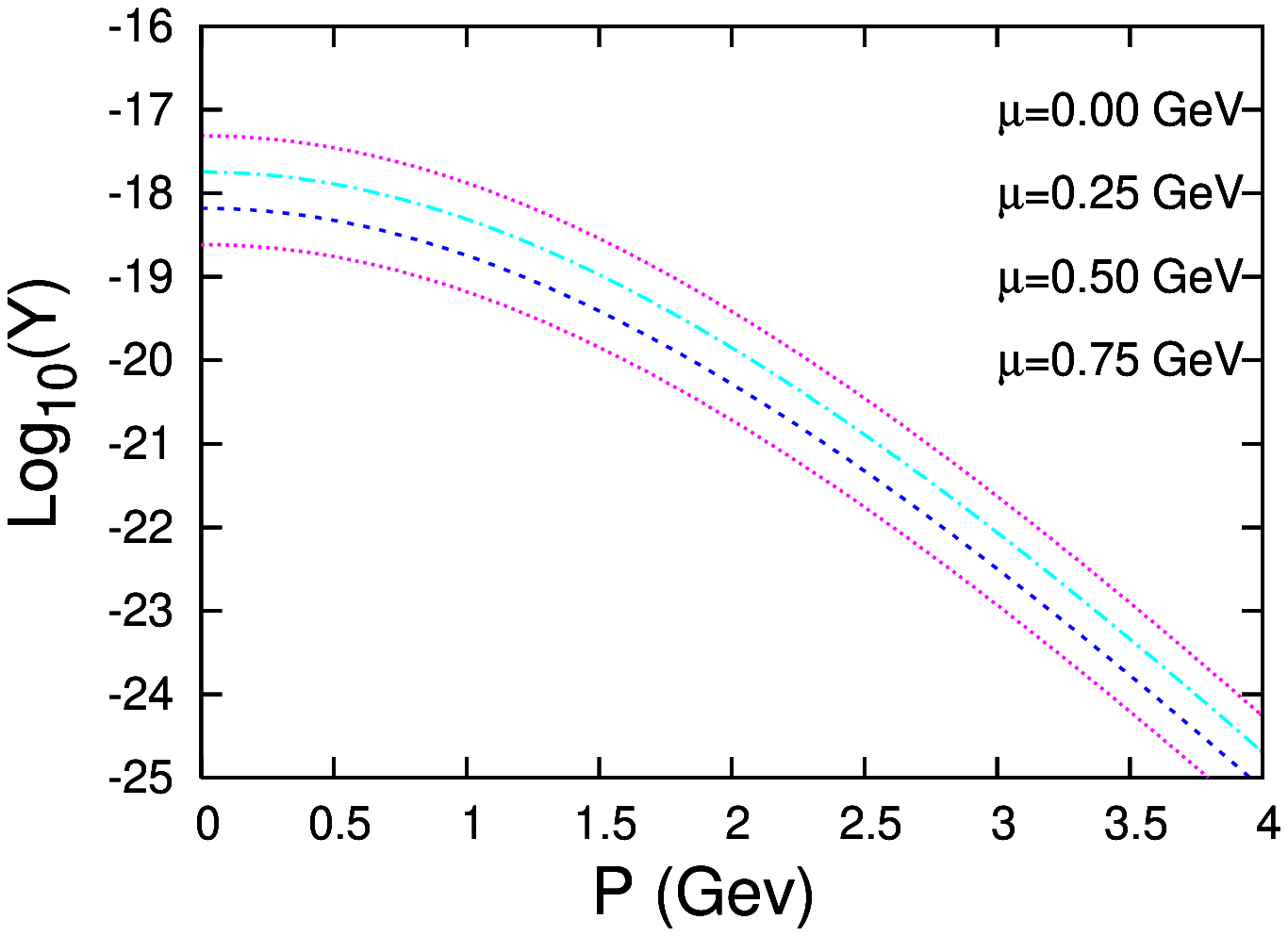}}}
\vspace*{0.5cm}
\caption[]{
The dilepton emission rate,$Y=\frac{dR}{d^{4}P}$, log($Y$) at parameter $M=3.0 GeV, T=0.25 GeV,K_{q}=0.6 GeV$ for different values of$\mu$}
\label{scaling}
\end{figure}
%%%%%%%%%%%%%%%%%%%%%%%%%%%%%%%%%%%%%%%%%%%%%%%%%%%%%%%%%%%%%%%%%%%%%%%%
\begin{figure}
\resizebox*{3.1in}{3.1in}{\rotatebox{270}{\includegraphics{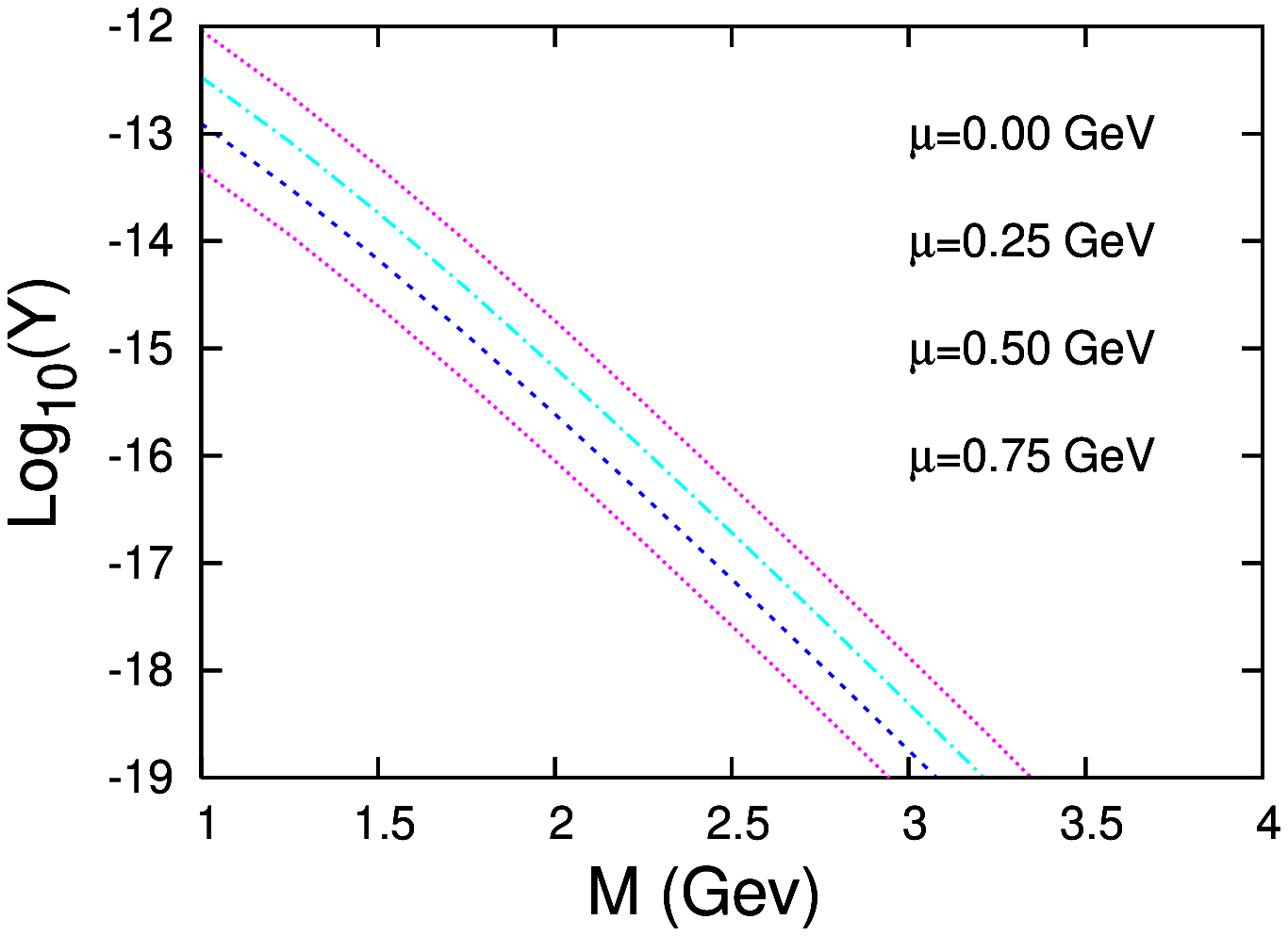}}}
\vspace*{0.5cm}
\caption[]{
The dilepton emission rate, $Y=\frac{dR}{d^{4}P}$, log($Y$) at parameter $P=1.0 GeV, T=0.25 GeV,K_{q}=0.6 GeV$ for diffent values of$\mu$}
\label{scaling}
\end{figure}
%%%%%%%%%%%%%%%%%%%%%%%%%%%%%%%%%%%%%%%%%%%%%%%%%%%%%%%%%%%%%%%%%%%%%%%%%%%%%%%
\begin{figure}
\resizebox*{3.1in}{3.1in}{\rotatebox{270}{\includegraphics{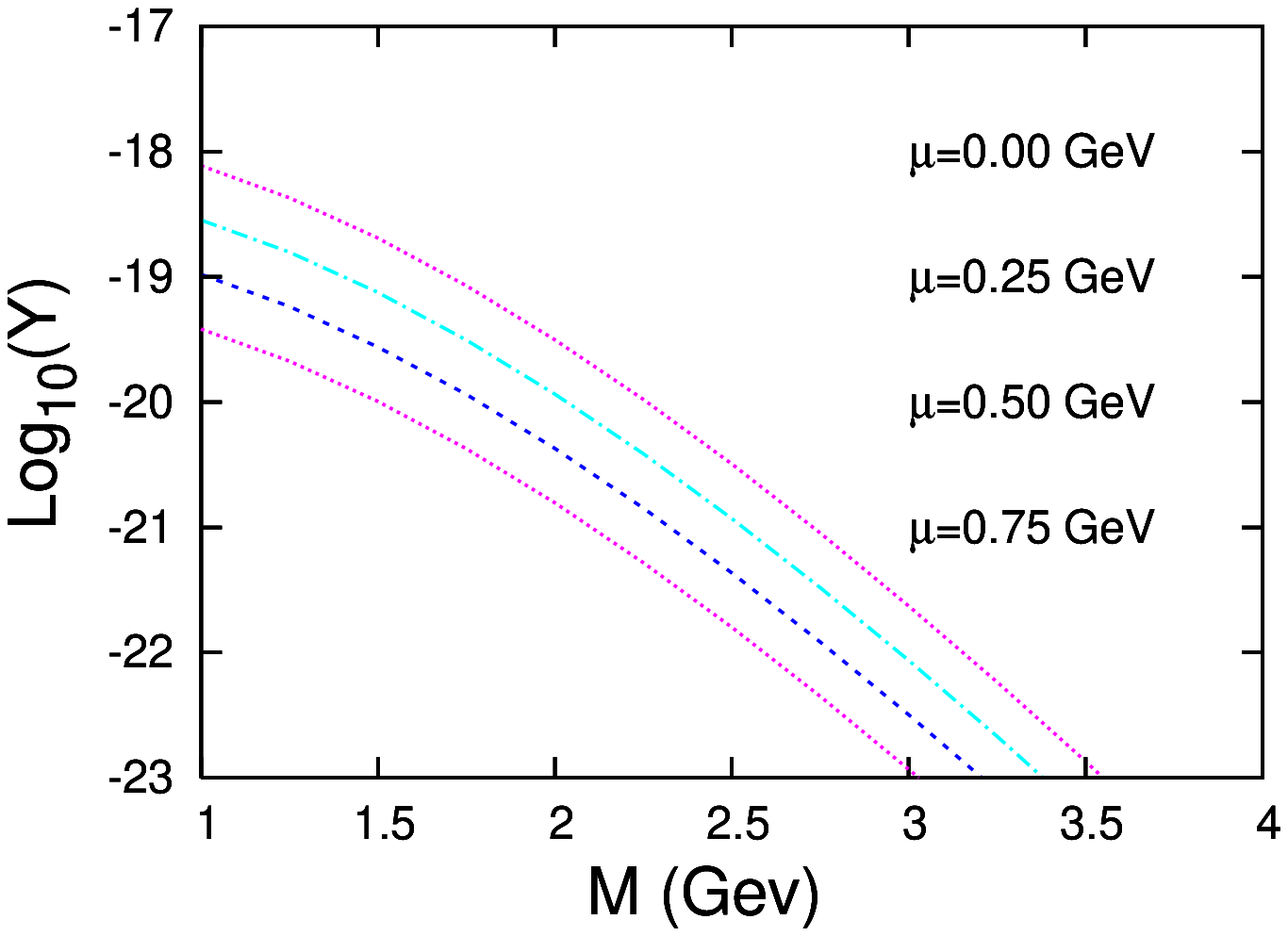}}}
\vspace*{0.5cm}
\caption[]{
The dilepton emission rate, $Y=\frac{dR}{d^{4}P}$, log($Y$) at parameter $P=3.0 GeV, T=0.25 GeV,K_{q}=0.6 GeV$ for diffent values of$\mu$}
\label{scaling}
\end{figure}
%%%%%%%%%%%%%%%%%%%%%%%%%%%%%%%%%%%%%%%%%%%%%%%%%%%%%%%%%%%%%%%%%%%%%%%%%%
\par
{\bf Results and Conclusions}: The results are 
shown in the figures.This result is consistant 
with the results produced by Ref[9] and that of Ref[11] for $\mu=0 $ 
which is free from the chemical potential. The cut off in the momentum 
distribution function is represented by the  threshold  
factor $\Theta(E-2K_{q})$ in the dilepton production. Now, in the figure 1 ,for small lepton pair mass $M=1.0 GeV$ and quark momentum cut off $ K_{q}=0.6 GeV$ we obtain 
 a sudden fall in the production rate of the dilepton and 
 at $ K_{q}=0 $ the rate  extends upto the level mark for $\mu=0$
which is shown by solid line .Again in the
 figure 2 we increase the lepton pair mass $M$ upto $3.0 GeV$ then 
the cut-off value of quark momentum $K_{q}=0.6 GeV$ does not play 
any role and sudden fall shown by figure 1 disappear.It implies  that the threshold factor 
$\Theta(E-2 K_{q})$ plays role mainly in small invariant lepton pair mass $M$
, not for large  $M$.Similarly it produces the variation of the 
emission rate with the lepton pair  mass in figure 3 for a specific 
transverse momentum $P_{T}=1.0 GeV$.From the analysis we again observed 
that cut-off in the momentum distribution function  plays role 
mainly in the small invariant  mass $M$ and for the large  value of $M$ 
it is not much effective .we obtain this  for the 
different values of $M=(1-4)GeV$ for  one fixed value of $P_{T}=1.0 GeV$.
Again we increase this transverse momemtum $P_{T}$ upto $3 GeV$ and 
the production rate obtained by this fixed $P_{T}$ value shows that 
at higher $P_{T}$ the rate is highly suppressed corresponding to the value of chemical potential.
Lastly we can conclude from this analysis that the rate of production 
  is decreasing function of the increasing in the chemical potential
in QGP which implies that in  the rich barion region the anti
quark population is strongly suppresssed for $\frac{\mu
}{T}>0$.The  result is  quiet similar with the results produced 
by Dimutru et. al.$[5]$for different values of $\mu/T=0,1,2,3$.So the effect of this evolution of the system at
finite baryon density on lepton production is very important.
\acknowledgements
  We are very thankful to Dr. R. Ramanathan for constructive 
suggestions and discussions.

\end{document}